\documentclass[12pt]{article}
\usepackage{latexsym,amsmath}
\input amssym

\begin{document}
\renewcommand{\theequation}{\thesection.\arabic{equation}}
\def\prg#1{\medskip{\bf #1}}
\def\lra{\leftrightarrow}        \def\Ra{\Rightarrow}
\def\nin{\noindent}              \def\pd{\partial}
\def\dis{\displaystyle}          \def\dfrac{\dis\frac}
\def\grl{{GR$_\Lambda$}}         \def\vsm{\vspace{-9pt}}
\def\Lra{{\Leftrightarrow}}      \def\ads3{AdS$_3$}
\def\cs{{\scriptscriptstyle \rm CS}}
\def\bull{\raise.25ex\hbox{\vrule height.8ex width.8ex}}
\def\Leff{\hbox{$\mit\L_{\hspace{.6pt}\rm eff}$}}
\def\Tr{\hbox{\rm Tr\hspace{1pt}}}
\def\bF{{\bar F}}                \def\ric{{(Ric)}}
\def\tmgl{\hbox{TMG$_\Lambda$}}
\def\mb#1{\mbox{\boldmath{$#1$}}}

\def\bA{{\bar A}}     \def\bx{{\bar x}}     \def\bT{{\bar T}}
\def\bH{{\bar H}}     \def\bL{{\bar L}}     \def\bB{{\bar B}}
\def\hO{{\hat O}}     \def\hG{{\hat G}}     \def\tG{{\tilde G}}
\def\cL{{\cal L}}     \def\cM{{\cal M }}    \def\cE{{\cal E}}
\def\cA{{\cal A}}     \def\cI{{\cal I}}     \def\cC{{\cal C}}
\def\cF{{\cal F}}     \def\hcF{\hat{\cF}}   \def\bcF{{\bar\cF}}
\def\cH{{\cal H}}     \def\hcH{\hat{\cH}}   \def\bcH{{\bar\cH}}
\def\cK{{\cal K}}     \def\hcK{\hat{\cK}}   \def\bcK{{\bar\cK}}
\def\hcT{\hat{\cT}}   \def\hH{{\hat H}}     \def\tH{{\tilde H}}
\def\cO{{\cal O}}     \def\hcO{\hat{\cal O}} \def\tR{{\tilde R}}
\def\cB{{\cal B}}     \def\bV{{\bar V}}     \def\heps{\hat\epsilon}
\def\cT{{\cal T}}     \def\bcT{{\bar\cT}}
\def\hcO{{\hat\cO}}   \def\hcH{\hat\cH}     \def\bL{{\bar L}}
\def\bcH{\bar\cH}     \def\bcK{\bar\cK}     \def\bD{{\bar D}}
\def\bG{{\bar G}}     \def\bd{\bar\delta}   \def\D{{\Delta}}
\def\bu{{\bar u}}     \def\bv{{\bar v}}     \def\bw{{\bar w}}

\def\G{\Gamma}        \def\S{\Sigma}        \def\L{{\mit\Lambda}}
\def\a{\alpha}        \def\b{\beta}         \def\g{\gamma}
\def\d{\delta}        \def\m{\mu}           \def\n{\nu}
\def\th{\theta}       \def\k{\kappa}        \def\l{\lambda}
\def\vphi{\varphi}    \def\ve{\varepsilon}  \def\p{\pi}
\def\r{\rho}          \def\Om{\Omega}       \def\om{\omega}
\def\s{\sigma}        \def\t{\tau}          \def\eps{\epsilon}
\def\ups{\upsilon}    \def\tom{{\tilde\om}} \def\bw{{\bar w}}
\def\nab{\nabla}      \def\tnab{{\tilde\nabla}}
\def\tcR{\tilde{\cal R}}
\def\Th{\Theta}       \def\cT{{\cal T}}     \def\cS{{\cal S}}
\def\nn{\nonumber}
\def\be{\begin{equation}}             \def\ee{\end{equation}}
\def\ba#1{\begin{array}{#1}}          \def\ea{\end{array}}
\def\bea{\begin{eqnarray} }           \def\eea{\end{eqnarray} }
\def\beann{\begin{eqnarray*} }        \def\eeann{\end{eqnarray*} }
\def\beal{\begin{eqalign}}            \def\eeal{\end{eqalign}}
\def\lab#1{\label{eq:#1}}             \def\eq#1{(\ref{eq:#1})}
\def\bsubeq{\begin{subequations}}     \def\esubeq{\end{subequations}}
\def\bitem{\begin{itemize}}           \def\eitem{\end{itemize}}
\title{Canonical structure of topologically massive gravity
with a cosmological constant}

\author{M. Blagojevi\'c and B. Cvetkovi\'c\footnote{
        Email addresses: {\tt mb@phy.bg.ac.yu,
                                cbranislav@phy.bg.ac.yu}} \\
Institute of Physics, P. O. Box 57, 11001 Belgrade, Serbia}
\date{}
\maketitle
\begin{abstract}
We study the canonical structure of three-dimensional topologically
massive gra\-vi\-ty with a cosmological constant, using the full
power of Dirac's method for constrained Hamiltonian systems. It is
found that the dimension of the physical phase space is two per
spacetime point, which corresponds to a single Lagrangian degree of
freedom. The analysis of the AdS asymptotic region reveals a
remarkable relation to 3D gravity with torsion: in the limit of
vanishing torsion, the conserved charges and asymptotic symmetries of
the two theories become identical.
\end{abstract}

\section{Introduction}
\setcounter{equation}{0}

Three-dimensional (3D) gravity, with or without a cosmological
constant $\L$, is a topological theory, in which there are no local
physical degrees of freedom \cite{1}. An interesting modification of
3D gravity is obtained by adding the gravitational Chern-Simons term.
General relativity with a Chern-Simons term is known as topologically
massive gravity (TMG), and in contrast to pure general relativity, it
is a dynamical theory with a local propagating degree of freedom, the
massive graviton \cite{2}. More generally, having in mind a rich
dynamical structure found in general relativity with a cosmological
constant \cite{3}, one expects that its extension by the
gravitational Chern-Simons term, denoted shortly as \tmgl, may
provide a new insight into the black hole dynamics and the asymptotic
structure of spacetime \cite{4}.

Both the gauge structure of a dynamical system and its physical
content are most clearly understood in the canonical formalism. The
constrained Hamiltonian analysis of the full \tmgl\ was carried out
recently in \cite{5,6,7} (for the case $\L=0$, see \cite{8}). The
treatment of the problem is characterized with complicated
calculational details, which might be a reason for significant
inconsistencies in the conclusions. Namely, Park \cite{5} found that
the number of degrees of freedom in configuration space is $N_c=3$
(one ``for each internal index"), Carlip \cite{6} obtained $N_c=1$,
while Grumiller et al. \cite{7} also found $N_c=1$, but in the chiral
version of the theory \cite{9}.

Our original motivation for studying \tmgl\ was to understand the
relation between 3D gravity and 3D gravity with torsion \cite{10,11},
and explore the influence of geometry on the gravitational dynamics.
After reading the literature, we learned that the constraint
structure of \tmgl\ has a rather controversial status \cite{5,6,7},
and we focused our attention on this issue. Our present study of the
structure of \tmgl\ is based on using the full power of Dirac's
canonical formalism \cite{12}, and it leads to the conclusion
$N_c=1$. The consistency of our results is checked by comparing with
the Lagrangian formalism, and by constructing the canonical gauge
generator. As a byproduct of our analysis, we are now able to
critically understand the results presented in the literature
\cite{5,6,7}. First, we discovered some errors in Park's
calculations, which is why his result for $N_c$ is {\it not
correct\/}. Second, although the values of $N_c$ obtained by Carlip
and by Grumiller et al. are correct, some aspects of the
corresponding derivations are {\it not satisfying\/}: they are based
on introducing an extra constraint by appealing to the Lagrangian
formalism, but the effect of this procedure on the overall constraint
structure of the theory remains unclear. Our systematic canonical
analysis gives a definitive support to the result $N_c=1$.

After clarifying the constraint structure of \tmgl, we extended our
analysis to the AdS asymptotic domain. Our study of the subject leads
to a remarkable relation between \tmgl\ and 3D gravity with torsion
\cite{11}: for a specific choice of parameters which ensures that the
torsion vanishes on shell, the conserved charges (energy and angular
momentum) and asymptotic symmetries of these two theories {\it
coincide\/}. This conclusion looks quite natural since it involves,
on shell, the Riemannian sector of 3D gravity with torsion. Another
interesting aspect of this relation is that it involves two theories
with substantially different dynamical contents: 3D gravity with
torsion is a topological theory, while \tmgl\ has one propagating
degree of freedom.

The paper is organized as follows. In section 2, we give a brief
account of the basic dynamical features of \tmgl\ in the Lagrangian
formalism. In sections 3 and 4, we apply Dirac's method for
constrained dynamical systems \cite{12} to make a complete analysis
of the constraint structure of \tmgl, which leads to $N_c=1$. In
section 5, we construct a convenient reduced phase space and use it
to make a comparison with Carlip's work \cite{6}. The construction of
the canonical gauge generator in section 6 confirms the consistency
of the previous analysis of constraints. Then, in section 7, we begin
the examination of the asymptotic structure of the theory by
introducing the AdS asymptotic conditions, which leads to a deep
relation between the asymptotic structures of \tmgl\ and 3D gravity
with torsion \cite{11}. The full content of this relation is
clarified in section 8, devoted to the canonical realization of the
asymptotic symmetry: we find the form of the surface term necessary
to make the canonical generator well-defined, calculate the conserved
charges and identify the central charges of the canonical algebra.
Finally, section 9 is devoted to concluding remarks, while appendices
contain some technical details.

Our conventions are given by the following rules: the Latin indices
refer to the local Lorentz frame, the Greek indices refer to the
coordinate frame;  the middle alphabet letters
$(i,j,k,...;\m,\n,\l,...)$ run over 0,1,2, the first letters of the
Greek alphabet $(\a,\b,\g,...)$ run over 1,2; the metric components in
the local Lorentz frame are $\eta_{ij}=(+,-,-)$; totally antisymmetric
tensor $\ve^{ijk}$ and the related tensor density $\ve^{\m\n\r}$ are
both normalized as $\ve^{012}=1$.

\section{The Lagrangian dynamics} 
\setcounter{equation}{0}

Topologically massive gravity with a cosmological constant is
formulated as a gravitational theory in Riemannian spacetime. Instead
of using the standard Riemannian formalism, with an action defined in
terms of the metric, we find it more convenient to use the triad
field and the spin connection as fundamental dynamical variables.
Such an approach can be naturally described in the framework of
Poincar\'e gauge theory \cite{13}, where basic gravitational
variables are the triad field $b^i$ and the Lorentz connection
$A^{ij}=-A^{ji}$ (1-forms), and the corresponding field strengths are
the torsion $T^i$ and the curvature $R^{ij}$ (2-forms). After
introducing the notation $A^{ij}=:-\ve^{ij}{_k}\om^k$ and
$R^{ij}=:-\ve^{ij}{_k}R^k$, we have:
$$
T^i=db^i+\ve^i{}_{jk}\om^j\wedge b^k\, ,\qquad
R^i=d\om^i+\frac{1}{2}\,\ve^i{}_{jk}\om^j\wedge\om^k\, .
$$

The antisymmetry of $A^{ij}$ ensures that the underlying geometric
structure corresponds to Riemann-Cartan geometry, in which $b^i$ is
an orthonormal coframe, $g:=\eta_{ij}b^i\otimes b^j$ is the metric of
spacetime, $\om^i$ is the Cartan connection, and $T^i,R^i$ are the
torsion and the Cartan curvature, respectively. For $T_i=0$, this
geometry reduces to Riemannian. In what follows, we will omit the
wedge product sign $\wedge$ for simplicity.

\prg{Field equations.} The Lagrangian of \tmgl\ is defined by
\be
L=2ab^i R_i-\frac{\L}{3}\,\ve_{ijk}b^i b^jb^k\,
+a\m^{-1}L_\cs(\om)+\l^i T_i\, ,                           \lab{2.1}
\ee
where $a=1/16\pi G$, $L_\cs(\om)=\om^id\om_i
+\frac{1}{3}\ve_{ijk}\om^i\om^j\om^k$ is the Chern-Simons Lagrangian
for the Lorentz connection, $\l^i$ (1-form) is the Lagrange
multiplier that ensures $T_i=0$.

The variation of the action $I=\int L$ with
respect to $b^i,\om^i$ and $\l^i$, yields the gravitational field
equations:
\bsubeq\lab{2.2}
\bea
&&2aR_i-\L\ve_{ijk}b^jb^k+\nab\l_i=0\, ,                   \lab{2.2a}\\
&&2aT_i+2a\m^{-1}R_i+\ve_{imn}\l^mb^n=0\, ,                \lab{2.2b}\\
&&T_i=0\, ,                                                \lab{2.2c}
\eea
\esubeq
where $\nab\l_i=d\l_i+\ve_{ijk}\om^j\l^k$ is the covariant derivative
of $\l_i$. With $T_i=0$, the second equation yields a simple solution
for $\l_m$:
$$
\l_m=2a\m^{-1}L_m\, ,\qquad
  L_m:=\left(\ric_{mn}-\frac{1}{4}\eta_{mn}R\right)b^n\, ,
$$
where $\ric_{mn}=-\ve^{kl}{_m}R_{kln}$, $R=-\ve^{ijk}R_{ijk}$. After
that, the first equation takes the form
\bsubeq
\be
2aR_i-\L\ve_{ijk}b^jb^k+2a\m^{-1}C_i=0\, ,                 \lab{2.3a}
\ee
where $C_i=\nab L_i$ is the Cotton 2-form. The expansion in the basis
$\heps_k=\frac{1}{2}\ve_{kmn}b^mb^n$, given by $R_i=G^k{}_i\heps_k$,
$C_i=C^k{}_i\heps_k$, yields the standard component form of the above
equation:
\be
aG_{ij}-\L\eta_{ij}+a\m^{-1}C_{ij}=0\, ,                   \lab{2.3b}
\ee
\esubeq
where $G_{ij}$ is the Einstein tensor, and $C_{ij}=\ve_i{}^{mn}\nab_m
L_{nj}$ the Cotton tensor.

For later convenience, we display here two simple consequences of the
field equations:
\be
\l_{mn}-\l_{nm}=0\, ,\qquad \m\l+3\L=0\, ,                 \lab{2.4}
\ee
where $\l=\l^n{_n}$.

\prg{Gauge symmetries.} By construction, gauge symmetries of the
theory \eq{2.1} are local translations and local Lorentz rotations,
parametrized by $\xi^\m$ and $\ve^{ij}=:-\ve^{ij}{_k}\th^k$. In local
coordinates $x^\m$, we have $b^i=b^i{_\m}dx^\m$,
$\om^i=\om^i{_\m}dx^\m$, $\l^i=\l^i{_\m}dx^\m$, and local Poincar\'e
transformations take the form:
 \bea
\d_P b^i{_\m}&=&-\ve^i{}_{jk}b^j{}_{\m}\th^k
   -(\pd_\m\xi^\r)b^i{_\r}-\xi^\r\pd_\r b^i{}_\m \, ,      \nn\\
\d_P\om^i{_\m}&=&-\nabla_\m\th^i-(\pd_\m\xi^\r)\om^i{_\r}
   -\xi^\r\pd_\r\om^i{}_\m \, ,                            \nn\\
\d_P\l^i{_\m}&=&-\ve^i{}_{jk}\l^j{}_{\m}\th^k
   -(\pd_\m\xi^\r)\l^i{_\r}-\xi^\r\pd_\r \l^i{}_\m \, .    \lab{2.5}
\eea

\prg{The BTZ black hole.} The BTZ black hole \cite{14}, a well-known
solution of the standard 3D gravity in the AdS sector (with
$\L=-1/\ell^2$), is a trivial solution of \tmgl, since the related
Cotton tensor identically vanishes.

In the Schwartzschield-like coordinates $x^\m=(t,r,\vphi)$, the BTZ
black hole solution is defined in terms of the lapse and shift
functions, respectively:
$$
N^2=\left(-8Gm+\frac{r^2}{\ell^2}+\frac{16G^2J^2}{r^2}\right)\, ,
  \qquad N_\vphi=\frac{4GJ}{r^2}\, .
$$
The triad field has the simple diagonal form
\bsubeq\lab{2.6}
\be
b^0=Ndt\, ,\qquad b^1=N^{-1}dr\, ,\qquad
b^2=r\left(d\vphi+N_\vphi dt\right)\, ,                    \lab{2.6a}
\ee
the connection reads
\be
\om^0=-Nd\vphi\, ,\qquad
  \om^1=N^{-1}N_\vphi dr\, ,\qquad
  \om^2= -\frac{r}{\ell^2}dt-N_\vphi rd\vphi\, ,           \lab{2.6b}
\ee
and the Lagrange multiplier is expressed in terms of the triad field
as
\be
\l^i=\frac{a}{\m\ell^2}b^i\, .                             \lab{2.6c}
\ee
\esubeq

Maximally symmetric solution of \tmgl, the AdS solution with isometry
group $SO(2,2)$, is formally obtained from \eq{2.6} by the replacements
$8mG=-1$, $J=0$.

\section{Hamiltonian and constraints}
\setcounter{equation}{0}

In order to get a deeper insight into dynamical structure of \tmgl,
we focus our attention on its canonical content \cite{12}. In local
coordinates $x^\m$, the component form of the Lagrangian density
reads:
\bea
\cL&=&\ve^{\m\n\r}\Bigl[ab^i{_\m}R_{i\n\r}
     -\frac{\L}{3}\ve_{ijk}b^i{_\m}b^j{_\n}b^k{_\r}        \nn\\
     &&+a\m^{-1}\Bigl(\om^i{_\m}\pd_\n\om_{i\r}
         +\frac{1}{3}\ve_{ijk}\om^i{_\m}\om^j{_\n}\om^k{_\r}\Bigr)
     +\frac{1}{2}\l^i{_\m}T_{i\n\r}\Bigr]\, .              \nn
\eea

\prg{1.} Introducing the canonical momenta $(\pi_i{^\m},\Pi_i{^\m},
p_i{^\m})$ corresponding to the Lagrangian variables
$(b^i{_\m},\om^i{_\m},\l^i{_\m})$, we find the primary constraints:
\bea
&&\phi_i{^0}:=\pi_i{^0}\approx 0\,,\qquad\,\,
  \phi_i{^\a}:=\pi_i{^\a}-\ve^{0\a\b}\l_{i\b}\approx 0\, , \nn\\
&&\Phi_i{^0}:=\Pi_i{^0}\approx 0\, ,\qquad
  \Phi_i{^\a}:=\Pi_i{^\a}
       -a\ve^{0\a\b}(2b_{i\b}+\m^{-1}\om_{i\b})\approx 0\,.\nn\\
&&p_i{^\m}\approx 0\,.                                     \lab{3.1}
\eea
The canonical Hamiltonian has the form:
\bea
&&\cH_c= b^i{}_0\cH_i+\om^i{}_0\cK_i+\l^i{_0}\cT^i+\pd_\a D^\a\,,\nn\\
&&\cH_i=-\ve^{0\a\b}\left(aR_{i\a\b}
        -\L\ve_{ijk}b^j{}_\a b^k{}_\b+\nabla_\a\l_{i\b}\right)\,,\nn\\
&&\cK_i=-\ve^{0\a\b}\left(aT_{i\a\b}+a\m^{-1}R_{i\a\b}
        +\ve_{ijk}b^j{}_\a \l^k{}_\b\right) \, ,           \nn\\
&&\cT_i=-\frac{1}{2}\ve^{0\a\b}T_{i\a\b}\,,                \nn\\
&& D^\a=\ve^{0\a\b}\left[ a\om^i{}_0\left( 2b_{i\b}
        +\m^{-1}\om_{i\b}\right)+b^i{}_0 \l_{i\b}\right]\, .\nn
\eea
The basic Poisson brackets (PBs) are displayed in Appendix A.

\prg{2.} Going over to the total Hamiltonian,
\be
\cH_T=b^i{}_0\cH_i+\om^i{}_0\cK_i+\l^i{}_0\cT_i
   +u^i{}_\m\phi_i{}^\m+v^i{}_\m\Phi_i{}^\m
   +w^i{}_\m p_i{^\m}+\pd_\a D^\a\, ,                      \lab{3.3}
\ee
we find that the consistency conditions of the primary constraints
$\pi_i{}^0$, $\Pi_i{}^0$ and $p_i{}^0$ yield the secondary
constraints:
\bsubeq
\be
\cH_i\approx0, \qquad \cK_i\approx0 \,, \qquad \cT_i\approx 0\,.
\ee

The consistency of the remaining primary constraints $\phi_i{}^\a$,
$\Phi_i{}^\a$ and $p_i{}^\a$ leads to the determination of the
multipliers $u^i{}_\b$, $v^i{}_\b$ and $w^i{}_\b$. Denoting the
determined multipliers by a bar, we have:
\bea
&&2a(\bv_{i\b}-\nabla_\b\om_{i0})+\bw_{i\b}+\ve_{ijk}\om^j{_0}\l^k{_\b}
  -\nabla_\b\l_{i0}-2\L\ve_{ijk}b^j{_0}b^k{_\b}=0\, ,      \nn\\
&&2a\m^{-1}(\bv_{i\b}-\nabla_\b\om_{i0})
  +\ve_{ijk}(b^j{_0}\l^k{_\b}-b^j{_\b}\l^k{_0})=0\, ,      \nn\\
&&\bu_{i\b}+\ve_{ijk}\om^j{_0}b^k{_\b}-\nabla_\b b_{i0}=0\,.
\eea
\esubeq
Using the Hamiltonian equations of motion $\dot b^i{}_\b=\bu^i{}_\b$,
$\dot\om^i{}_\b=\bv^i{}_\b$ and $\dot \l^i{}_\b=\bw^i{}_\b$, these
relations reduce to the $(0,\b)$ components of the Lagrangian field
equations \eq{2.2}.

The substitution of the determined multipliers into \eq{3.3} yields
the modified form of the total Hamiltonian:
\bea
&&\cH_T=\hcH_T+\pd_\a\bD^\a\, ,                            \nn\\
&&\hcH_T=b^i{}_0\bcH_i+\om^i{}_0\bcK_i+\l^i{_0}\bcT_i
         +u^i{}_0\pi_i{}^0+v^i{}_0\Pi_i{}^0+w^i{_0}p_i{^0}\,,\nn
\eea
where
\bea
&&\bcH_i=\cH_i-\nabla_\b\phi_i{}^\b
  -\frac{\m}{2a}\ve_{ijk}\l^j{_\b}\Phi^{k\b}
  +\ve_{ijk}\left(2\L b^j{_\b}+\m\l^j{_\b}\right)p^{k\b}\,,\nn\\
&&\bcK_i=\cK_i-\ve_{ijk}b^j{}_\b\phi^{k\b}-\nabla_\b\Phi_i{}^\b
  -\ve_{ijk}\l^j{_\b}p^{k\b}\, ,                           \nn\\
&&\bcT_i=\cT_i
  -\frac{\m}{2a}\ve_{ijk} b^j{_\b}\Phi^{k\b}-\nabla_\b p_i{^\b}
  +\m\ve_{ijk}b^j{}_\b p^{k\b}\, ,                         \nn\\
&&\bD^\a=D^\a+b^i{}_0\phi_i{}^\a+\om^i{}_0\Phi_i{}^\a
  +\l^i{_0}p_i{^\a}\, .                                    \nn
\eea

\prg{3.} The consistency conditions of the secondary constraints
read:
\bea\lab{3.5}
&&\{\bcH_i,H_T\}\approx -\frac{\m}{2a}\ve^{0\a\b}\Bigl[b_{i0}\l_{\a\b}
 -\l_{i\a}(\l_{0\b}-\l_{\b 0})\Bigr]=:X_i\, ,              \nn\\
&&\{\bcT_i,H_T\}\approx \frac{\m}{2a}\ve^{0\a\b}\Bigl[b_{i0}\l_{\a\b}
 -b_{i\a}(\l_{0\b}-\l_{\b 0})\Bigr]=:Y_i\, ,               \nn\\
&&\{\bcK_i,H_T\}\approx 0\, ,
\eea
where $\l_{\m\n}=b^k{_\m}\l_{k\n}$. This result contains an important
difference with respect to the one obtained by Park, Eq. (14) in
\cite{5}, which consists in the presence of the $\l_{\a\b}$ terms. To
count the number of independent tertiary constraints, one notes that
$Y_i\approx 0$ is equivalent to
\bsubeq\lab{3.6}
\bea
&&\th_{0\b}:=\l_{0\b}-\l_{\b 0}\approx 0\, ,               \\
&&\th_{\a\b}:=\l_{\a\b}-\l_{\b\a}\approx 0\, ,
\eea
\esubeq
which, in turn, ensures $X_i\approx 0$. Thus, we have only {\it three
independent\/} tertiary constraints, $\th_{0\b}$ and $\th_{\a\b}$,
which are the canonical equivalents of the Lagrangian relations
\eq{2.4}$_1$.

\prg{4.} The consistency of $\th_{\a\b}$ yields
\bea
\{\th_{\a\b},H_T\}&=&
  \bu^k{_\a}\l_{k\b}+b^k{_\a}\bw_{k\b}-(\a\lra\b)
  \approx-2b\ve_{0\a\b}\Bigl(3\L+\m\l\Bigr)\approx 0\, .   \nn
\eea
Thus, we have a new, quartic constraint:
\be
\Psi=3\L+\m\l\approx 0\, .                                 \lab{3.8}
\ee
The quartic constraint is a canonical equivalent of the Lagrangian
relation \eq{2.4}$_2$.

To interpret the consistency condition for $\th_{0\b}$, we introduce
the notation
\be
\pi_i{^0}{}':=\pi_i{^0}+\l_i{^k}p_k{^0}\, ,\qquad
w^i{_0}{}':=w^i{_0}-u^k{_0}\l_k{^i}\, .
\ee
The $(\pi_i{^0},p_i{^0})$ piece of the Hamiltonian can be written in
the form
\bea
u^i{_0}\pi_i{^0}+w^i{_0}p_i{^0}&=&u^i{_0}\pi_i{^0}{}'
  +w^i{_0}{}'p_i{^0} \, .                                  \nn
\eea
The consistency of $\th_{0\b}$ imposes a condition on the two
components $w'_{\b 0}=w_{m0}{}'b^m{_\b}$ of $w_i{^0}{}'$:
\bea
&&\{\th_{0\b},H_T\}=
  (b^i{_0}\bw_{i\b}-\l^i{_0}\bu_{i\b})
  -(w^m{_0}-u^k{_0}\l_k{^m})b_{m\b}\approx 0\, ,           \nn\\
&&\bw'_{\b 0}=b^i{_0}\bw_{i\b}-\l^i{_0}\bu_{i\b}\, .       \lab{3.8a}
\eea

\prg{5.} Finally, the consistency requirement on $\Psi$ determines
$w'_{00}=w'_{m0}b^m{_0}$:
\bea
&&\{\Psi,H_T\}=g^{00}\bw'_{00}+g^{\b 0}\bw'_{\b 0}
               +h^{i\b}(\bw_{i\b}-\l_i{^k}\bu_{k\b})\approx 0\,,\nn\\
&&g^{00}\bw'_{00}=(\l^{i\b}+\l^i{_0}g^{0\b})\bu_{i\b}-
                (h^{i\b}+b^i{_0}g^{0\b})\bw_{i\b}\, .      \lab{3.8b}
\eea
This completes the consistency procedure.

The final form of the total Hamiltonian can be written as
\bea
\hcH_T&=&\bar\cH_T
  +u^i{_0}\pi_i{^0}{}'+v^i{_0}\Pi_i{^0}\, ,                \\
\bar\cH_T&:=&b^i{_0}\bcH_i+\om^i{_0}\bcK_i+\l^i{_0}\bcT_i
             +\bw_{\b 0}'p^{\b 0}+\bw_{00}'p^{00}\, .      \nn
\eea

\section{Classification of constraints} 
\setcounter{equation}{0}

Among the primary constraints, those that appear in $\cH_T$ with
arbitrary multipliers are first class (FC):
\bsubeq
\be
\pi_i{^0}{}',\Pi_i{^0}=\mbox{FC}\, ,
\ee
while the remaining ones are second class.

Going to the secondary constraints, we use the following simple
theorem:
\bitem
\item[$\bull$] If $\phi$ is a FC constraint, then $\{\phi,H_T\}$
is also a FC constraint.
\eitem
The proof relies on using the Jacoby identity. The theorem implies
that the secondary constraints $\hcH_i:=-\{\pi_i{^0}{}',H_T\}$ and
$\hcK_i:=-\{\Pi_i{^0},H_T\}$ are FC. After a lengthy but
straightforward calculation, we obtain:
\bea
&&\hcH_i=\bcH'_i
          +h_i{^\r}(\nab_\r \l_{jk})b^k{_0}p^{j0}\, ,      \nn\\
&&\hcK_i=\bcK_i-\ve_{ijk}(\l^j{_0}p^{k0}-b^j{_0}\l^k{_n}p^{n0})\,,
\eea
\esubeq
where $\bcH':=\bcH_i+\l_i{^k}\bcT_k$. In deriving the above form of
$\hcH_i$, we used the weak equality
$$
2\L\ve_{inm}+\m
  \left(\ve_{ink}\l^k{_m}-\ve_{imk}\l^k{_n}\right)\approx
  h_n{^\m}h_m{^\n}(\nab_\m\l_{i\n}-\nab_\n\l_{i\m})\, ,
$$
where time derivatives are expressed in terms of the determined
multipliers.

The PB algebra between the FC constraints ($\hcH_i,\hcK_j$) is
calculated in Appendix A. The total Hamiltonian can be expressed in
terms of the FC constraints as follows:
\be
\hcH_T=b^i{_0}\hcH_i+\om^i{_0}\hcK_i
  +u^i{_0}\pi_i{^0}{}'+v^i{_0}\Pi_i{^0}-\th_{0\b}h^{n\b}\hcT_n\, ,
\ee
where the last term is an ignorable square of constraints, with
\be
\hcT_n:=-\{\pi_n{^0},H_T\}=\bcT_n-b_{n0}\pd_\b p^{\b 0}
  -(\nabla_\b b_{n0})p^{\b 0}
  -\m\ve_{njk}b^j{_0}b^k{_\b}p^{\b 0}\, .                  \nn
\ee

The complete classification of constraints is summarized in Table 1.
\begin{center}
\doublerulesep 1.8pt
\begin{tabular}{lll}
\multicolumn{3}{l}{\hspace{16pt}Table 1. Classification of contraints}\\
                                                       \hline\hline
\rule{0pt}{12pt}
&~First class \phantom{x}&~Second class \phantom{x}   \\
                                                      \hline
\rule[-1pt]{0pt}{15pt}
\phantom{x}Primary &~$\p_i{^0}{}',\Pi_i{^0}$
            &~$\phi_i{^\a},\Phi_i{}^{\a},p_i{^\a},p_i{^0}$  \\
                                                      \hline
\rule[-1pt]{0pt}{15pt}
\phantom{x}Secondary\phantom{x} &~$\hcH_i,\hcK_i$
            &~$\bcT_i$                                \\
                                                      \hline
\rule[-1pt]{0pt}{15pt}
\phantom{x}Tertiary\phantom{x}
                   &~ &~$\th_{0\b},\th_{\a\b}$        \\
                                                      \hline
\rule[-1pt]{0pt}{15pt}
\phantom{x}Quartic\phantom{x}
                   &~ &~$\Psi$                        \\
                                                      \hline\hline
\end{tabular}
\end{center}
The content of Table 1 related to the second class constraints needs
additional explanation. We begin by noting that the primary
constraints $(\phi_i{^\a},\Phi_i{}^{\a},p_i{^\a},p_i{^0})$ are of the
second class, as the related multipliers in $H_T$ are determined. The
second class nature of the remaining constraints
$(\bcT_i,\th_{0\b},\th_{\a\b},\Psi)$ can be verified by analyzing
their PB algebra. There is, however, a much simpler argument based on
the counting of dynamical degrees of freedom, as explained bellow.

When the classification of constraints is complete, the number of
dynamical degrees of freedom in the phase space is given by the
formula:
\be
N^*=2N-2N_1-N_2\, ,                                        \nn
\ee
where $N$ is the number of Lagrangian dynamical variables, $N_1$ is
the number of FC, and $N_2$ the number of second class constraints.
According to our results, we have $N=27$, $N_1=12$ and $N_2=28$, the
dimension of the phase space is $N^*=2$, and the theory exhibits one
local Lagrangian degree of freedom, the topologically massive
graviton \cite{2,8}.

The argument that supports the classification displayed in Table 1
goes as follows. If at least two constraints in the set
$(\bcT_i,\th_{0\b},\th_{\a\b},\Psi)$ were FC, then $N^*$ would be
negative. This is, however, {\it not possible\/}, hence, all the
constraints $(\bcT_i,\th_{0\b},\th_{\a\b},\Psi)$ are of the second
class. A more technical argument on this point is given in the next
section.

\section{The reduced phase space} 
\setcounter{equation}{0}

The canonical analysis of \tmgl\ developed so far is based on using
the full phase space with coordinates
$(b^i{_\m},\om^i{_\m},\l^i{_\m};\pi_i{^\m},\Pi_i{^\m},p_i{^\m})$.
Now, we wish to examine what happens when we go to the reduced phase
space formalism, in which the PBs are replaced by the Dirac brackets
(DB) \cite{12}.

We begin by noting that we have two sets of FC constraints,
$\pi_i{^0}{}'$ and $\Pi_i{^0}$, hence we are free to impose two sets
of gauge conditions. A simple and natural choice is to fix the form
of the corresponding unphysical variables, $b^i{_0}$ and $\om^i{_0}$.
This can be done, for instance, by demanding their forms to coincide
with the black hole solution. After that, we can construct the
corresponding DBs and eliminate the variables
$(b^i{_0},\pi_i{^0}{}')$ and $(\om^i{_0},\Pi_i{^0})$ from the theory;
the DBs of the remaining variables remain unchanged. Note that
similar arguments cannot be applied to the pair $(p_i{^0},\l^i{_0})$,
since $p_i{^0}$ is not a FC constraint.

Next, we use the second class constraints
$X_A:=(\phi_i{^\a},\Phi_i{^\a},p_i{^\a})$ to eliminate the remaining
momenta $(\pi_i{^\a},\Pi_i{^\a},p_i{^\a})$. After that, the structure
of the reduced phase space $R_1$ with canonical coordinates
$(b^i{_\a},\om^i{_\a},\l^i{_\a};\l^i{_0},p_i{^0})$ is determined by
the DBs
\bea
&&\{b^i{_\a},b^j{_\b}\}^*_1=0\, ,\qquad
  \{b^i{_\a},\om^j{_\b}\}^*_1=0\, ,\qquad
  \{b^i{_\a},\l^j{_\b}\}^*_1=\ve_{0\a\b}\eta^{ij}\d\, ,    \nn\\
&&\{\om^i{_\a},\om^j{_\b}\}^*_1=\frac{\m}{2a}\ve_{0\a\b}\eta^{ij}\d\,,
  \qquad \{\om^i{_\a},\l^j{_\b}\}^*_1=
                 -\m\ve_{0\a\b}\eta^{ij}\d\, ,             \nn\\
&&\{\l^i{_\a},\l^j{_\b}\}^*_1=
  2a\m\ve_{0\a\b}\eta^{ij}\d\, ,                           \lab{5.1}
\eea
plus those involving $\l^i{_0}$ and $p_i{^0}$ (Appendix B).

Finally, we introduce the reduced phase space $R_2$, defined by the 6
second class constraints $Y_A:=(\th_{0\b},\Psi,p^{\a 0}, p_0{^0})$.
The constraints $Y_A$ can be used to eliminate $\l^i{_0}$ and
$p_i{^0}$ from $R_1$, whereupon the reduced phase space $R_2$ is
described by the canonical coordinates
$(b^i{_\a},\om^i{_\a},\l^i{_\a})$. Using the iterative property of
DBs, the influence of $Y_A$ on the form of DBs is described by the
matrix $\D_2$, with $(\D_2)_{AB}=\{Y_A,Y_B\}^*_1\,$ (Appendix B).
Explicit calculation shows that the form of the new DBs is defined by
the following simple rule:
\bitem
\item[\bull] The new DBs in $R_2$ are the same as those in Eq. \eq{5.1}.
\eitem

The classification of constraints in $R_2$ is displayed in Table 2.
\begin{center}
\doublerulesep 1.8pt
\begin{tabular}{lll}
\multicolumn{3}{l}{\hspace{16pt}Table 2. Classification of
                                constraints in $R_2$} \\
                                                      \hline\hline
\rule{0pt}{12pt}
&~First class \phantom{x}&~Second class \phantom{x}   \\
                                                      \hline
\rule[-1pt]{0pt}{15pt}

\rule[-1pt]{0pt}{15pt}
\phantom{x}Secondary\phantom{x} &~$\cH'_i,\cK_i$
            &~$\cT_i$                                 \\
                                                      \hline
\rule[-1pt]{0pt}{15pt}
\phantom{x}Tertiary\phantom{x}
                   &~ &~$\th_{\a\b}$                  \\
                                                      \hline\hline
\end{tabular}
\end{center}
The number of the phase space variables is $3\times 6=18$, there are
$6$ first class and $4$ second class constraints, and the number of
physical degrees of freedom is the same as before, $N^*=18-2\times
6-4=2$, as it should.

Treating $(b^i{_0},\om^i{_0},\l^i{_0})$ as Lagrange multipliers,
Carlip worked from the very beginning in the reduced phase with
canonical coordinates $(b^i{_\a},\om^i{_\a},\l^i{_\a})$ \cite{6}.
To compare his construction with our $R_2$, we replace the
variables $\om^i$ and $\l^i$ by $A^i=\om^i+\m b^i$ and
$\b^i=\l^i-a\m b^i$, respectively. The resulting non-trivial DBs
are: \be \{A^i{_\a},A^j{_\b}\}_2^*=
             \frac{\m}{2a}\ve_{0\a\b}\eta^{ij}\d\, ,\qquad
\{b^i{_\a},\b^j{_\b}\}_2^*=\ve_{0\a\b}\eta^{ij}\d\, , \lab{5.2}
\ee in complete agreement with Eq. (3.2) in \cite{6} (in units
$a=1$). Hence, $R_2$ coincides with Carlip's construction of the
phase space.

At this stage, one can check the second class nature of
$Z_A=(\cT_i,\th_{\a\b})$ directly from the form of their DBs:
\bea
&&\{\cT_i,\cT_j\}^*_2=\frac{\m}{2a}\ve^{0\a\b}b_{i\a}b_{j\b}\,,\nn\\
&&\{\cT_i,\th_{\a\b}\}^*_2=\nabla_\b(b_{i\a}\d)-\nabla_\a(b_{i\b}\d)
  +2\m\ve_{imn}b^m{_\a}b^n{_\b}\d\, ,                      \nn\\
&&\{\th_{\a\b},\th_{\g\d}\}^*_2=0\, .
\eea
Indeed, as shown in \cite{6}, the matrix
$(\D_3)_{AB}=\{Z_A,Z_B\}^*_2$ is invertible.

\section{Gauge generator} 
\setcounter{equation}{0}

After completing the Hamiltonian analysis, we now wish to construct
the canonical gauge generator \cite{15}. Starting from the primary FC
constraints $\pi_i{^0}{}'$ and $\Pi_i{^0}$, one finds:
\bea
&&G[\t]=\dot\t^i\pi_i{^0}{}'
  +\t^i\left[\hcH^i-\ve_{ijk}\om^j{_0}\pi^k{^0}'
  +\frac \m{2a}\left(\ve_{imn}\l_j{^n}
  -\ve_{jmn}\l_i{^n}\right)b^j{_0}\Pi^{m0}\right]\, ,      \nn\\
&&G[\s]=\dot\s^i \Pi_i{^0}+\s^i\left(\hcK^i
  -\ve_{ijk}\om^j{_0}\Pi^k{_0}-\ve_{ijk}b^j{_0}\pi^{k0}{}'\right)\,.
\eea
The complete gauge generator has the form $G=G[\t]+G[\s]$, its action
on the fields is defined by the PB operation $\d_0\phi=\{\phi,G\}$,
but the resulting gauge transformations do not have the Poincar\'e
form \eq{2.5}. The standard Poincar\'e content of the gauge
transformations is obtained by introducing the new parameters
\cite{11}
$$
\t^i=-\xi^\r b^i{_\r}\, ,\qquad \s^i=-\th^i-\xi^\r \om^i{_\r}\, .
$$
Expressed in terms of these parameters (and after neglecting some
trivial terms, quadratic in the constraints), the gauge generator
takes the form:
\bea
G&=&-G_1-G_2\,,\nn\\
G_1&=&\dot\xi^\r\left(b^i{}_\r\pi_i{}^0
       +\l^i{_\r}p_i{}^0+\om^i{}_\r\Pi_i{}^0\right)        \nn\\
&&+\xi^\r\left[b^i{}_\r\bcH_i+\l^i{_\r}\bcT_i
  +\om^i{}_\r\bcK_i+(\pd_\r b^i_0)\pi_i{}^0
  +(\pd_\r\l^i{_0})p_i{^0}+(\pd_\r\om^i{}_0)\Pi^i{}_0\right]\,,\nn\\
G_2&=&\dot{\th^i}\Pi_i{}^0+\th^i\left[\bcK_i
  -\ve_{ijk}\left(b^j{}_0\pi^{k0}+\l^j{}_0p^{k0}
  +\om^j{}_0\Pi^{k0}\right)\right]\, .                     \lab{6.2}
\eea
Looking at the related gauge transformations, we find a complete
agreement with the Poincar\'e gauge transformations \eq{2.5} \emph{on
shell}.

\section{Asymptotic conditions} 
\setcounter{equation}{0}

Asymptotic conditions imposed on dynamical variables determine the
form of asymptotic symmetries, and consequently, they are closely
related to the gravitational conservation laws. In this section, we
focus our attention to the AdS sector of the theory, characterized by
the negative value of the cosmological constant:
$$
\frac{\L}{a}=:-\frac 1{\ell^2}\, .
$$

\prg{AdS asymptotics.} The AdS asymptotic conditions are introduced
by demanding that (a) the asymptotic configurations include the black
hole solution \eq{2.6}, and (b) they are invariant under the action
of the AdS group $SO(2,2)$. Following the procedure defined in 3D
gravity with torsion \cite{11}, we find the asymptotic form for the
triad field:
\bsubeq\lab{7.1}
\be
b^i{_\m}= \left( \ba{ccc}
       \dis\frac{r}{\ell}+\cO_1   & O_4  & O_1  \\
       \cO_2 & \dis\frac{\ell}{r}+\cO_3  & O_2  \\
       \cO_1 & \cO_4                     & r+\cO_1
                 \ea
          \right)\, ,                                      \lab{7.1a}
\ee
and for the connection:
\be
\om^i{_\m}=\left( \ba{ccc}
  \cO_1 & \cO_2  & \dis -\frac{r}{\ell}+\cO_1 \\
  \cO_2 & \cO_3  & \cO_2                      \\
  \dis-\frac{r}{\ell^2}+\cO_1 & \cO_2 & \cO_1
                  \ea
           \right)\, .                                     \lab{7.1b}
\ee
In \tmgl, we have one more Lagrangian variable, the Lagrange
multiplier $\l^i$. Since $\l^i$ for the black hole solution satisfies
\eq{2.6c}, we define its asymptotic behavior by the relation:
\be
\l^i{_\m}=\frac{a}{\m\ell^2}b^i{_\m}+\hcO\, ,              \lab{7.1c}
\ee
\esubeq
where $\hcO$ denotes terms with arbitrarily fast asymptotic decrease.

At this stage, by comparing \eq{7.1a} and \eq{7.1b} with the
asymptotic conditions in 3D gravity with torsion, see section 4 in
\cite{11}, we are led to an important observation:
\bitem
\item[(A1)] The asymptotic form of $b^i{_\m}$ and $\om^i{_\m}$ in
\tmgl\ is the same as in 3D gravity with torsion in the limit when
the torsion vanishes on shell.
\eitem
Looking at the field equations of 3D gravity with torsion displayed
in Appendix C, one finds that the condition of vanishing torsion
takes the form $p=0$, where $p$ is a combination of the coupling
constants. The origin of this property may be traced back to the form
of the BTZ black hole \eq{2.6}. As we shall see in the next section,
(A1) lies at the root of a remarkable correspondence between the
asymptotic structures of \tmgl\ and 3D gravity with torsion.

\prg{Asymptotic parameters.} Having chosen the asymptotic conditions
in the form \eq{7.1}, we now wish to find the subset of gauge
transformations that respect these conditions. As a first consequence
of (A1), we conclude that the parameters of the restricted gauge
transformations have the same form as in 3D gravity with torsion
\cite{11}:
\bea
&&\xi^0=\ell\left[ T
  +\frac{1}{2}\left(\frac{\pd^2 T}{\pd t^2}\right)
              \frac{\ell^4}{r^2}\right] +\cO_4\, ,\qquad
  \xi^1=-\ell\left(\frac{\pd T}{\pd t}\right)r+\cO_1\, ,   \nn\\
&&\xi^2=S-\frac{1}{2}\left(\frac{\pd^2 S}{\pd\vphi^2}\right)
              \frac{\ell^2}{r^2}+\cO_4\, ,                 \lab{7.2}
\eea
and similarly for $\th^i$. Here, the functions $T(t,\vphi)$ and
$S(t,\vphi)$ are determined by the conditions
$$
T^-=T^-(x^-)\,,\quad T^+=T^+(x^+)\, ,
$$
where $T^\mp=T\mp S$ and  $x^\mp=x^0/\ell\mp x^2$. After expressing
$T^\mp$ in terms of the Fourier modes and introducing the notation
$\d_P(T^\mp=e^{inx^\mp})=:\ell_n^\mp$, the asymptotic commutator
algebra takes the familiar form of two independent Virasoro algebras
without central charges:
$$
i[\ell_n^-,\ell^-_m]=(n-m)\ell_{n+m}^-\, ,\qquad
i[\ell_n^+,\ell^+_m]=(n-m)\ell_{n+m}^+\, .
$$
The asymptotic symmetry of spacetime, defined by the parameters
$T^\mp$, coincides with the conformal symmetry.

\prg{Asymptotics of the phase space.} In order to extend the
asymptotic conditions \eq{7.1} to the canonical level, one should
determine an appropriate asymptotic behavior of the momentum variables.
This step is based on the following general principle: the expressions
than vanish on shell should have an arbitrary fast asymptotic decrease,
as no solutions of the field equations are thereby lost. By applying
this principle to the primary constraints \eq{3.1}, one finds the
asymptotic behavior of all the momentum variables.

\section{Canonical realization of the asymptotic symmetry} 
\setcounter{equation}{0}

In this section, we study the influence of the adopted asymptotic
conditions on the canonical structure of \tmgl: we construct the
improved gauge generators, examine their canonical algebra and prove
the conservation laws. As a consequence of (A1), all these
characteristics are naturally related the the corresponding results
in 3D gravity with torsion.

\subsection{Surface terms} 

The canonical generator acts on dynamical variables via the PB
operation, hence, it should have well-defined functional derivatives.
In order to ensure this property, we have to improve the form  of $G$
by adding a suitable surface term $\G$, such that $\tG=G+\G$ is a
well-defined canonical generator. In this process, the asymptotic
conditions play a crucial role \cite{16,11}.

Following the same calculational technique as in \cite{11}, we find
that the improved canonical generator takes the form
\bsubeq\lab{8.1}
\bea
&&\tG=G+\G\, ,                                             \nn\\
&&\G:=-\oint df_\a\left(\xi^0\cE^\a+\xi^2\cM^\a\right)
     =-\int_0^{2\pi}d\vphi\left(\ell T\cE^1+S\cM^1\right)\,,
\eea
where
\bea
&&\cE^\a=2\ve^{0\a\b}\left(a\om^0{_\b}+\frac{a}{2\m\ell^2}b^0{_\b}
  +\frac{1}{2}\l^0{_\b}+\frac a{\ell}b^2{_\b}
  +\frac{a}{\m\ell}\om^2{_\b}\right)b^0{_0}\, ,             \nn\\
&&\cM^\a=-2\ve^{0\a\b}\left(a\om^2{_\b}+\frac{a}{2\m\ell^2}b^2{_\b}
  +\frac 12\l^2{_\b}+\frac a{\ell}b^0{_\b}
  +\frac{a}{\m\ell}\om^0{_\b}\right)b^2{_2}\, .
\eea
\esubeq

Now, we can use the asymptotic relation \eq{7.1c} for $\l^i{_\m}$ and
compare the value of the surface term $\G$ with the
corresponding expression for 3D gravity with torsion, displayed in
Appendix C, with the following conclusion:
\bitem
\item[(A2)] The value of the surface integral $\G$ in the AdS
sector of \tmgl\ coincides with the corresponding value in 3D gravity
with torsion, in the limit of vanishing torsion.
\eitem
This conclusion is a natural consequence of (A1).

\subsection{Conserved charges}

The values of the surface terms, calculated for $\xi^0=1$ and
$\xi^2=1$, define the energy and angular momentum of the system,
respectively:
\be
E=\int_0^{2\pi}d\vphi\,\cE^1\, ,\qquad
M=\int_0^{2\pi}d\vphi\,\cM^1\, .
\ee
In particular, the energy and angular momentum for the BTZ black hole
\eq{2.6} are:
\be
E=m-\frac{J}{\m\ell^2}\, ,\qquad
M=J-\frac{m}{\m}\, .
\ee
In agreement with (A2), these BTZ charges are seen to coincide with
the corresponding expressions in 3D gravity with torsion, in the
limit of vanishing torsion, see Appendix C.

\subsection{Canonical algebra}

Using the notation $\tG_{(i)}:=\tG[T_i^+,T_i^-]$, the main theorem of
\cite{17} states that the canonical algebra of the improved
generators has the general form:
\bsubeq\lab{8.5}
\be
\left\{\tG_{(2)},\tG_{(1)}\right\}= \tG_{(3)}+C_{(3)}\, ,
\ee
where $C_{(3)}$ is the central term. To calculate $C_{(3)}$, we note
that
$$
\{\tG_{(2)},\tG_{(1)}\}\approx \d_{(1)}\G_{(2)}
                       \approx \G_{(3)}+C_{(3)}\, .
$$
The calculation of $\d_{(1)}\G_{(2)}$ is based on the asymptotic
transformation laws of the ener\-gy/angular momentum densities
$\cE_\mp=(\ell\cE^1\mp\cM^1)/2$:
$$
\d\cE_\mp=-T^\mp\pd_\mp\cE_\mp -2(\pd_\mp T^\mp)\cE_\mp
  +a\ell\left(1\pm\frac{1}{\ell\m}\right)\pd_\mp^3T^\mp\, ,
$$
and it leads to
\bea
&&C_{(3)}=C_-[T^-]+C_+[T^+]\, ,                            \nn\\
&&C_\mp[T^\mp]:=-a\ell\left(1\pm\frac{1}{\ell\m}\right)
   \int_0^{2\pi}d\vphi(\pd_\mp^3 T_1^\mp)T_2^\mp\, .
\eea
\esubeq
Introducing the Fourier modes for the improved generator,
$L^\mp_n=-\tG[T^\mp=e^{inx^\mp}]$, the canonical algebra \eq{8.5}
takes the form of two independent Virasoro algebras with different
central charges:
\be
c^\mp=24\pi a\ell\left(1\pm\frac{1}{\ell\m}\right)
     =\frac{3\ell}{2G}\left(1\pm\frac{1}{\ell\m}\right)\, .
\ee
A direct comparison with Appendix C implies that the central charges
of \tmgl\ have the same values as in the $p=0$ limit of 3D gravity
with torsion, which is, again, a consequence of the general
correspondence (A2).

Once we have the central charges, we can use Cardy's formula to
obtain the black hole entropy \cite{4,18}:
\be
S=\frac{2\pi r_+}{4G}-\frac{2\pi r_-}{4G\mu\ell}\,,
\ee
where $r_+$ and $r_-$ (the radii of the outer and inner black hole
horizon, respectively) are related to the black hole parameters $m$
and $J$ by $r_+^2+r_-^2=8Gm\ell^2$, $r_+r_-=4GJ\ell$. The form of the entropy is in
agreement with the first law of black hole thermodynamics.

\section{Concluding remarks} 
\setcounter{equation}{0}

In this paper, we studied \tmgl\ as a constrained dynamical system
\cite{11}. Our approach is based on using the triad field $b^i$ and
the spin connection $\om^i$ as independent dynamical variables, while
the Lagrange multiplier $\l^i$ is introduced to ensure the vanishing
of torsion. Our goal was twofold: first, to obtain and classify the
constraints and deduce the dimension of the physical phase space
$N^*$, and second, to examine the asymptotic structure of \tmgl\ and
compare it with the corresponding features of 3D gravity with
torsion.

(1) With regard to the first goal, we found $N^*=2$, which means that
the number of Lagrangian degrees of freedom is $N_c=1$.

Since Park \cite{5} used the same formalism, we can easily compare
his results with ours. Park's consistency conditions for $\bcH_i$ and
$\bcT_i$ in Eq. (14) of \cite{5} are not correctly calculated, as one
can see by comparing with our Eqs. \eq{3.5} and \eq{3.6}. As a
consequence, Park missed the tertiary constraint $\th_{\a\b}$.
Without $\th_{\a\b}$, he was not able to find the quartic constraint
$\Psi$, given in our Eq. \eq{3.8}. Moreover, one can directly
conclude that Park's classification of constraints is not correct.
Indeed, if $\Psi_a^0$ and $\bcK_a$ ($\Pi_i{^0}$ and $\bcK_i$ in our
notation) were the only FC constraints as claimed in \cite{5}, we
would not be able to construct the complete Poincar\'e gauge
generator, but only its Lorentz piece. Consequently, $N_c=3$ is not
the correct result.

As we mentioned in section 5, Carlip treated
$(b^i{_0},\om^i{_0},\l^i{_0})$ as Lagrange multipliers, and he worked
in the phase space equivalent to our $R_2$ \cite{6}. After
identifying the secondary constraints (as defined in Table 2), he
relied on the Lagrangian formalism to justify the introduction of an
{\it extra\/} constraint $\D$ (in section 4). Adding a constraint in
this way is a serious step, which might influence dynamical content
of the original theory. To prevent that, one needs a consistency
control of the procedure which guarantees that the constraint
content of the theory remains unchanged with respect to the genuine
canonical treatment. In particular, one should clarify whether there
exist some other Lagrangian expressions, beside $\Delta$, that should
be also treated as constraints. We have not found a satisfying
analysis of these issues in \cite{6}. The extra constraint $\D$
essentially coincides with our $\th_{\a\b}$.

For negative $\L$, one can define the chiral version of \tmgl\ by
demanding that one of the two central charges vanishes, $\m\ell\mp
1=0$. Li et al. \cite{9} argued that, while \tmgl\ for generic $\m$
is unstable, the chiral version of the theory might be consistent.
Grumiller et al. \cite{7} studied the case $\m\ell=1$ in a reduced
phase space formalism, which is simmilar to (but not identical with)
the one used by Carlip. They found $N_c=1$, but again, only after
imposing the additional condition $\th_{\a\b}\approx 0$, whose
canonical status was not discussed. Our results imply that transition
to the chiral coupling does not have a critical influence on the form
of the PB algebra. Hence, we have $N_c=1$ also for the chiral
coupling.

(2) As a consistency check of our analysis of constraints, we used
the PB algebra to construct the canonical generator of Poincar\'e
gauge transformations. The form of this generator is improved by
adding suitable surface terms, and used to examine the AdS asymptotic
structure of \tmgl. The result of this analysis leads to a remarkable
conclusion: the conserved charges and asymptotic symmetries of \tmgl\
are the same as in 3D gravity with torsion, in the limit of vanishing
torsion. It is interesting to note that we have here two theories
with substantially different local properties (3D gravity with
torsion is a topological theory, while \tmgl\ has one propagating
degree of freedom), but still, they have classically identical
asymptotic structures.

\appendix
\section*{Acknowledgements} 
We wish to thank Steve Carlip and Mu-In Park for useful discussions.
This work was supported by the Serbian Science Foundation under grant
141036.

\section{The algebra of constraints} 
\setcounter{equation}{0}

In this appendix, we display the most important PBs that facilitate
the evaluation of the consistency requirements. Starting from the
basic relations $\{b^i{_\m},\pi_j{^\n}\}=
                 \d^i_j\d^\n_\m\d(\mbox{\boldmath{$x-x'$}})$ etc.,
we find the PBs between the primary constraints,
\bea
&&\{\phi_i{}^\a,\Phi_j{}^\b\}=-2a\ve^{0\a\b}\eta_{ij}\d\,,\qquad
  \{\phi_i{^\a},p_j^{\b}\}=-\ve^{0\a\b}\eta_{ij}\d\, ,     \nn\\
&&\{\Phi_i{}^\a,\Phi_j{}^\b\}
  = -2a\m^{-1}\ve^{0\a\b}\eta_{ij}\d\, ,                   \nn
\eea
between the primary and secondary constraints,
\bea
&&\{\phi_i{^\a},\bcH_j\}=2\L\ve_{ijk}p^{k\a}\d\, ,\qquad
  \{\phi_i{^\a},\bcK_j\}=-\ve_{ijk}\phi^{k\a}\d\, ,        \nn\\
&&\{\phi_i{^\a},\bcT_j\}=\frac{\m}{2a}\ve_{ijk}\left(-\Phi^{k\a}
                         +2ap^{k\a}\right)\d\,,            \nn\\
&&\{\Phi_i{^\a},\bcH_j\}=-\ve_{ijk}\phi^{k\a}\, ,\quad
  \{\Phi_i{^\a},\bcK_j\}=-\ve_{ijk}\Phi^{k\a}\d\, ,        \nn\\
&&\{\Phi_i{^\a},\bcT_j\}=-\ve_{ijk}p^{k\a}\d\, ,           \nn\\
&&\{p_i{^\a},\bcH_j\}=\frac{\m}{2a}\ve_{ijk}\left(-\Phi^{k\a}
                        +2ap^{k\a}\right)\d\, ,\qquad
  \{p_i{^\a},\bcK_j\}=-\ve_{ijk}p^{k\a}\d\, ,              \nn
\eea
and the PBs between the secondary constraints,
\bea
\{\bcH_i,\bcH_j\}&=&2\L\ve_{ijk}\bcT^k\d
    +\frac{\m}{2a}\ve^{0\a\b}\l_{i\a}\l_{j\b}\d            \nn\\
&&+\frac{\m}{2a}\ve_{ijk}\ve_{mn}{^k}\l^m{_\b}\left[
    2a\left(2ap^{n\b}-\Phi^{n\b}\right)
    +\phi^{n\b}\right]\d\, ,                               \nn\\
\{\bcH_i,\bcK_j\}&=&-\ve_{ijk}\bcH^k\d\, ,                 \nn\\
\{\bcH_i,\bcT_j\}&=&\frac{\m}{2a}\ve_{ijk}\left(-\bcK^k
                             +2a\bcT^k\right)\d
    -\frac{\m}{2a}\ve^{0\a\b}
           (\eta_{ij}\l_{\a\b}+\l_{i\a}b_{j\b})\d\, ,      \nn\\
&&-\frac{\m}{2a}\ve_{imk}\ve_{jn}{^k}b^m{_\b}\left[
    2a\left(2ap^{n\b}-\Phi^{n\b}\right)+\phi^{n\b}\right]
    +\frac{\m}{2a}\ve_{imk}\ve_{jn}{^k} p^{m\b}\l^n{_\b}\d\,,\nn\\
\{\bcK_i,\bcK_j\}&=&-\ve_{ijk}\bcK^k\d\, ,\qquad
    \{\bcK_i,\bcT_j\}=-\ve_{ijk}\bcT^k\d\, ,               \nn\\
\{\bcT_i,\bcT_j\}&=&\frac{\m}{2a}\ve^{0\a\b}b_{i\a}b_{j\b}\d
  +\frac{\m}{2a}\left(b_{i\b}p_j{^\b}-b_{j\b}p_i{^\b}\right)\d\,.\nn
\eea
Next, we calculate the PBs between $(\th_{0\b},\th_{\a\b})$ and the
secondary constraints:
\bea
&&\{\th_{0\b},\bcH_i\}=\nab'_\b(\l_i{_0}\d)
  +\ve_{imk}\Big(2\L b^m{_\b}+\m\l^m{_\b}\Big)b^k{_0}\d\,, \nn\\
&&\{\th_{0\b},\bcK_i\}=-\ve_{imk}\big(\l^m{_\b}b^k{_0}
  +\l^m{_0}b^k{_\b}\big)\d\, ,                             \nn\\
&&\{\th_{0\b},\bcT_i\}=-\nabla'_\b(b_{i0}\d)
  +\m\ve_{ijk}b^j{_\b}b^k{_0}\d\, ,                        \nn\\
&&\{\th_{\a\b},\bcH_i\}=-\nab'_\a(\l_{i\b}\d)
  -\ve_{ijk}b^j{_\a}(2\L b^k{_\b}
  +\m\l^k{_\b})\d-(\a\leftrightarrow\b)\, ,                \nn\\
&&\{\th_{\a\b},\bcK_i\}=0 \nn\\
&&\{\th_{\a\b},\bcT_i\}=-\nab'_\b(b_{i\a}\d)
  -\m\ve_{ijk}b^j{_\a}b^k{_\b}-(\a\leftrightarrow\b)\, ,   \nn
\eea
and between $\Psi$ and the secondary constraints:
\bea
&&\{\Psi,\bcH_i\}= \m\left[\nab'_\b(\l^\b{_i}\d)
  -\ve_{ijk}h^{j\a}(2\L b^k{_\a}+\m\l^k{_\a})\d\right]\, , \nn\\
&&\{\Psi,\bcK_i\}= -\m\ve_{ijk}\left(h^{j0}\l^k{_0}
  +b^j{_0}\l^0{_k}\right)\d\, ,                            \nn\\
&&\{\Psi,\bcT_i\}=\m\left[-\nab'_\b(h_i{^\b}\d)
  +\m\ve_{ijk}h^{j0}b^k{_0}\d\right]\, .                   \nn
\eea

Finally, we display the PBs among the secondary first class
constraints ($\hcH_i,\hcK_j$):
\bea
&&\{\hcH_i,\hcH_j\}=
  -\frac{\m}{2a}\ve_{ijk}\l^k{_m}\hcK^m\d\, ,              \nn\\
&&\{\hcH_i,\hcK_j\}=-\ve_{ijk}\hcH^k\d\, ,                 \nn\\
&&\{\hcK_i,\hcK_j\}=-\ve_{ijk}\hcK^k\d\, .                 \nn
\eea

\section{Dirac brackets} 
\setcounter{equation}{0}

The phase space $R_1$ is defined by the second class constraints
$X_A:=(\phi_i{^\a},\Phi_i{^\a},p_i{^\a})$. To construct the
corresponding DBs, we consider the $18\times 18$ matrix $\D_1$ with
matrix elements $(\D_1)_{AB}=\{X_A.X_B\}$:
\be
\D_1=\left( \ba{ccc}
    \{\phi_i{^\a},\phi_j{^\b}\}&\{\phi_i{^\a},\Phi_j{^\b}\}
                               &\{\phi_i{^\a},p_j{^\b}\}   \\
    \{\Phi_i{^\a},\phi_j{^\b}\}&\{\Phi_i{^\a},\Phi_j{^\b}\}
                               &\{\Phi_i{^\a},p_j{^\b}\}   \\
    \{p_i{^\a},\phi_j{^\b}\}&\{p_i{^\a},\Phi_j{^\b}\}
                            &\{p_i{^\a},p_j{^\b}\}
           \ea
     \right)\, .                                           \nn
\ee
The explicit form of $\D_1$ reads:
\be
\D_1(\mb{x},\mb{y})=
     \left( \ba{ccc}
                0& -2a   & -1     \\
              -2a& -2a\m^{-1}&  0 \\
               -1&  0    &  0
               \ea
     \right)\otimes\ve^{0\a\b}\eta_{ij}\d(\mb{x},\mb{y})\, .\nn
\ee
The matrix $\D_1$ is regular, and its inverse has the form
\be
\D_1^{-1}(\mb{y},\mb{z})=\frac{\m}{a}
     \left( \ba{ccc}
                       0 & 0             & a\m^{-1} \\
                       0 &\dis\frac{1}{2}& -a       \\
                     a\m^{-1}& a             & 2a^2
            \ea
     \right)\otimes\ve_{0\b\g}\eta^{jk}\d(\mb{y},\mb{z})\,.\nn
\ee
The matrix $\D_1^{-1}$ defines the DBs in the phase space $R_1$:
$$
\{\phi,\psi\}^*_1=\{\phi,\psi\}
                  -\{\phi,X_A\}(\D_1^{-1})^{AB}\{X_B,\psi\}\, .
$$
The
main part of the result is displayed in \eq{5.1}, while the remaining
non-trivial first-level DBs involving $\l^i{_0},p^{\b 0}$ and
$p_0{^0}$ are:
\bea
&&\{\l^i{_\a},p^{\b 0}\}_1^*=-\ve_{0\a\g}h^{i\b}p^{\g0}\d\,,\nn\\
&&\{\l^i{_0},p^{\b 0}\}_1^*=h^{i\b}\d\, ,\qquad
  \{\l^i{_0},p_0{^0}\}_1^*= b^i{_0}\d\, .                  \nn
\eea

The reduced phase space $R_2$ is obtained from $R_1$ by imposing the
additional second class constraints $Y_A:=(\th_{0\b},\Psi,p^{\a 0},
p_0{^0})$. The corresponding $6\times 6$ matrix $\D_2$ reads:
\be
\D_2:=\left(\ba{cccc}
\{\th_{0\a},\th_{0\b}\}_1^* & \{\th_{0\a},\Psi\}_1^* &
    \{\th_{0\a},p^{\b0}\}_1^* &\{\th_{0\a},p_0{^0}\}^* \\
\{\Psi,\th_{0\b}\}_1^* & \{\Psi,\Psi\}_1^* &
    \{\Psi,p^{\b0}\}_1^*&\{\Psi,p_0{^0}\}_1^*          \\
\{p^{\a0},\th_{0\b}\}_1^* & \{p^{\a0},\Psi\}_1^* &
    \{p^{\a0},p^{\b0}\}_1^*&\{p^{\a0},p_0{^0}\}_1^*    \\
\{p_0{^0},\th_{0\b}\}_1^* & \{p_0{^0},\Psi\}_1^* &
    \{p_0{^0},p^{\b0}\}_1^* & \{p_0{^0},p_0{^0}^*\}
\ea\right)\, .                                             \nn
\ee
The explicit form of $\D_2$ is:
$$
\D_2(\mb{x},\mb{y})=\left(\ba{cc}
                           B & A \\
                        -A^T & 0
                          \ea
                    \right)\d(\mb{x}-\mb{y})\, ,
$$
where
$$
A:=-\left(\ba{cc}
         \d_\a^\b & g_{0\a} \\
          -g^{0\b}& 1
         \ea
    \right)\, , \quad
B:=\left(\ba{cc}
  2\ve_{0\a\b}\left[a\m g_{00}-\l_{00}\right]&\d^0_\a \\
      \d^0_\b & 0
\ea
\right)\, .
$$
The inverse of $\D_2$ is given by
\bea
&&(\D_2)^{-1}(\mb{y},\mb{z})=
             \left(\ba{cc}
                  0 & -(A^T)^{-1}       \\
              A^{-1}& A^{-1}B(A^T)^{-1}
                  \ea
             \right)\d(\mb{y}-\mb{z})\, ,                  \nn\\
&&A^{-1}=\frac{1}{g_{00}g^{00}}
             \left(\ba{cc}
    -\d_\a^\b+\ve_{0\a\g}\ve^{0\b\eps}g^{0\g}g_{0\eps} & -g_{0\a} \\
                     g^{0\b} & 1
                   \ea
             \right)\, .                                   \nn
\eea
The DBs in $R_2$ are the same as those in \eq{5.1}.

\section{3D gravity with torsion in brief}
\setcounter{equation}{0}

Here, we give here a short review of some relevant features of the
topological Mielke-Baekler model \cite{10,11}. The model is defined
by the Lagrangian
$$
L=2ab^i R_i-\frac{\L}{3}\,\ve_{ijk}b^i b^j b^k\
    +\a_3L_\cs(\om)+\a_4 b^i T_i\, .
$$
In the non-degenerate sector with $\a_3\a_4-a^2\neq 0$, the
gravitational field equations have the form
$$
2T_i=p\ve_{ijk}b^jb^k\,,\qquad 2R_i=q\ve_{ijk}b^jb^k\, ,
$$
where
$$
p:=\frac{\a_3\L+\a_4 a}{\a_3\a_4-a^2}\, ,\qquad
q:=-\frac{(\a_4)^2+a\L}{\a_3\a_4-a^2}\, .
$$
The Riemannian piece of the Cartan curvature reads:
$$
2\tR_i=\Leff\ve_{ijk}b^jb^k\,,\qquad\Leff:=q-\frac{p^2}4\,.
$$
In the AdS sector of the theory, where the effective cosmological
constant $\Leff$ is negative, we have $\Leff=:-1/\ell^2$.

The surface term of the improved canonical generator reads:
\bea
&&\G:=-\int_0^{2\pi}d\vphi
         \left(\xi^0\cE^1+\xi^2\cM^1\right)\, ,            \lab{C1}\\
&&\cE^\a=
    2\ve^{0\a\b}\left[\left(a+\frac{\a_3p}{2}\right)\om^0{}_\b
   +\left(\a_4+\frac{ap}{2}\right)b^0{}_\b+\frac{a}{\ell}b^2{}_\b
   +\frac{\a_3}{\ell}\om^2{}_\b\right]b^0{}_0\, ,          \nn\\
&&\cM^\a=
   -2\ve^{0\a\b}\left[\left(a+\frac{\a_3p}{2}\right)\om^2{}_\b
   +\left(\a_4+\frac{ap}{2}\right)b^2{}_\b+\frac{a}{\ell}b^0{}_\b
   +\frac{\a_3}{\ell}\om^0{}_\b\right]b^2{}_2\, .          \nn
\eea

The values of the surface term for $\xi^0=1$ and $\xi^2=1$ define the
energy and angular momentum of the system, respectively. In
particular, the conserved charges for the BTZ black hole read:
\be
E= m+\frac{\a_3}{a}\left(\frac{pm}{2}-\frac{J}{\ell^2}\right)\,,
\qquad M= J+\frac{\a_3}{a}\left(\frac{pJ}{2}-m\right)\, .  \lab{C2}
\ee
The canonical algebra of the improved generators is characterized by
two different central charges:
\be
c^\mp=\frac{3\ell}{2G}
      +24\pi\a_3\left(\frac{p\ell}{2}\pm 1\right)\, .      \lab{C3}
\ee

According to the field equations, the vanishing of torsion can be
described by three equivalent conditions:
\be
p=0\, ,\qquad q=\frac\L a=-\frac1{\ell^2}\, ,\qquad
\a_4=\frac{\a_3}{\ell^2}\, .                               \lab{C4}
\ee
This case is of particular interest for comparison with \tmgl. Note
that the Chern-Simons coupling constant $\m$  in \tmgl\ is related to
$\a_3$ by $\a_3={a}/{\m}$.

For $p=0$, the treatment of the chiral limit of 3D gravity with
torsion demands an extension of the canonical analysis to the sector
$\a_3\a_4-a^2=0$.


\begin{thebibliography}{99} 

\bibitem{1}  S. Deser, R. Jackiw and  G 't Hooft,
  Three-dimensional Einstein gravity: dynamics of flat space,
  Ann. Phys. (N.Y) {\bf 152} (1984) 220;\\
  S. Deser and R. Jackiw, Three-dimensional cosmological gravity:
  dynamics of constant curvature,
  Ann. Phys. (N.Y) {\bf 153} (1984) 405;\\
  E. Martinec,  Soluble systems in Quantum Gravity,
  Phys. Rev. D {\bf 30} (1984) 1198.

\bibitem{2} S. Deser, R. Jackiw and S. Templeton,
  Three-Dimensional Massive Gauge Theories, Phys. Rev. Lett. {\bf 48}
  (1982) 975; Topologically Massive Gauge Theories,
  Ann. Phys. {\bf 140} (1982) 372.

\bibitem{3}   For a review of the subject and an extensive list of
  references, see: S. Carlip, Conformal Field Theory, (2+1)-dimensional
  Gravity, and the BTZ Black Hole, Class. Quant. Grav. {\bf 22} (2005)
  R85-R124.

\bibitem{4} P. Kraus and F. Larsen, Holographic gravitational anomalies,
  JHEP {\bf 0601} (2006) 022;\\
  S. N. Solodukhin, Holography with Gravitational Chern-Simons
  Term, Phys. Rev. D {\bf 74} (2006) 024015;\\
  K. Hotta, Y. Hyakutake, T. Kubota and H. Tanida,
  Brown-Henneaux's Canonical Approach to Topologically Massive
  Gravity, JHEP {\bf 0807} (2008) 066.

\bibitem{5} Mu-In Park, Constraint Dynamics and Gravitons in Three
  Dimensions, JHEP {\bf 0809} (2008) 084.

\bibitem{6} S. Carlip, The Constraint Algebra of Topologically Massive
  AdS Gravity, JHEP {\bf 0810} (2008) 078.

\bibitem{7} D. Grumiller, R. Jackiw and N. Johansson,
  Canonical analysis of cosmological topologically massive
  gravity at the chiral point, arXiv:0806.4185v1[hep-th],
  contribution to Wolfgang Kummer Memorial Volume.

\bibitem{8} S. Deser and X. Xiang, Canonical formulations of full nonlinear
  topologically massive gravity, Phys. Lett. B {\bf 263} (1991) 39.\\
  I. L. Buchbinder, S. L. Lyahovich and V. A. Krychin,
  Canonical quantization of topologically massive gravity,
  Class. Quant. Grav. {\bf 10} (1993) 2083.

\bibitem{9} W. Li, W. Song and A. Strominger, Chiral Gravity in
  Three Dimensions, JHEP {\bf 0804} (2008) 082;\\
  A. Strominger, A Simple Proof of the Chiral Gravity Conjecture,\\
  preprint arXiv:0808.0506v1[hep-th].

\bibitem{10}  E. W. Mielke, P. Baekler, Topological gauge model of
  gravity with torsion, Phys. Lett. A {\bf 156} (1991) 399;\\
  P. Baekler, E. W. Mielke, F. W. Hehl,
  Dynamical symmetries in topological 3D gravity with torsion,
  Nuovo Cim. B {\bf 107} (1992) 91.

\bibitem{11} M. Blagojevi\'c and B. Cvetkovi\'c, Canonical structure of
  3D gravity with torsion, in: {\it Progress in General Relativity and
  Quantum Cosmology\/}, vol. 2, ed. Ch. Benton (Nova Science Publishers,
  New York, 2006), pp. 103-123 (preprint gr-qc/0412134);
  Asymptotic charges in 3D gravity with torsion,
  Journal of Physics: Conf. Series {\bf 33} (2006) 248.

\bibitem{12} P. A. M. Dirac, {\it Lectures on Quantum Mechanics\/}
  (Belfer Graduate School of Science, Yeshiva University, New York,
  1964).

\bibitem{13} M. Blagojevi\'c, {\it Gravitation and gauge symmetries\/}
  (IoP Publishing, Bristol, 2002);\\
  T. Ort\'in, {\it Gravity and strings\/}, (Cambridge University Press,
  Cambridge, 2004).

\bibitem{14} M. Ba\~nados, C. Teitelboim and J. Zanelli, The Black Hole
  in Three-Dimensional Spacetime, Phys. Rev. Lett. {\bf 16}, 1849
  (1993);\\
  M. Ba\~nados, M. Henneaux, C. Teitelboim and J. Zanelli, Geometry of
  2+1 Black Hole, Phys. Rev. D {\bf 48}, 1506 (1993).

\bibitem{15} L. Castellani, Symmetries of constrained Hamiltonian
  systems, Ann. Phys. (N.Y.) {\bf 143}, 357 (1982).

\bibitem{16} T. Regge and C. Teitelboim, Role of surface integrals in
  the Hamiltonian formulation of general relativity, Ann. Phys. (N.Y.)
  {\bf 88} (1974) 286.

\bibitem{17} J. D. Brown and M. Henneaux, On the Poisson bracket of
  differentiable generators in classical field theory, J. Math. Phys.
  {\bf 27} (1986) 489.

\bibitem{18} M. Blagojevi\'c and B. Cvetkovi\'c, Black hole entropy from
  the boundary conformal structure in 3D gravity with torsion,
  JHEP {\bf 10} (2006) 005.

\end{thebibliography}
\end{document}